\documentclass[]{jfm}

\usepackage{graphicx}
\usepackage{newtxtext}
\usepackage{newtxmath}
\usepackage{natbib}
\usepackage{hyperref}
\hypersetup{
    colorlinks = true,
    urlcolor   = blue,
    citecolor  = black,
}

\newcommand{\RomanNumeralCaps}[1]
\linenumbers

\newcommand{\bfu}{\boldsymbol{u}} 
\newcommand{\bfv}{\boldsymbol{v}} 
\newcommand{\bfp}{\boldsymbol{p}} 
\newcommand{\bfx}{\boldsymbol{x}} 
 
\newcommand{\bfz}{\boldsymbol{z}} 
\newcommand{\bfr}{\boldsymbol{r}} 

\newcommand{\phihat}{\skew3\hat{\boldsymbol{\phi}}}
\newcommand{\iden}{\mathsfbi{I}}
\newcommand{\intd}{\,\mathrm{d}}
\newcommand{\bfn}{\boldsymbol{n}} 
\newcommand{\bfk}{\boldsymbol{k}} 
\newcommand{\bff}{\boldsymbol{F}} 
\newcommand{\bfkhat}{\hat{\boldsymbol{k}}} 
\newcommand{\pd}[2]{\frac{\partial #1}{\partial #2}}
\newcommand{\lsd}{\ell}
\newcommand{\bffhat}{\skew5\hat{\boldsymbol{f}}}

\title{Stability of a dispersion of elongated particles embedded in a viscous membrane}

\author{Harishankar Manikantan
  \corresp{\email{hmanikantan@ucdavis.edu}}}

\affiliation{Department of Chemical Engineering, University of California, Davis, CA 95616, USA}

\begin{document}
\maketitle

\begin{abstract}
We develop a mean-field model to examine the stability of a `quasi-2D suspension' of elongated particles embedded within a viscous membrane. This geometry represents several biological and synthetic settings, and we reveal mechanisms by which the anisotropic mobility of particles interacts with long-ranged viscous membrane hydrodynamics. We first show that a system of slender rod-like particles driven by a constant force is unstable to perturbations in concentration -- much like sedimentation in analogous 3D suspensions -- so long as membrane viscous stresses dominate. However, increasing the contribution of viscous stresses from the surrounding 3D fluid(s) suppresses such an instability. We then tie this result to the hydrodynamic disturbances generated by each particle in the plane of the membrane and show that enhancing subphase viscous contributions generates extensional fields that orient neighboring particles in a manner that draws them apart. The balance of flux of particles aggregating versus separating then leads to a wave number selection in the mean-field model. 
\end{abstract}

%\begin{keywords}
%%Authors should not enter keywords
%\end{keywords}

%{\bf MSC Codes }  {\it(Optional)} Please enter your MSC Codes here

\section{Introduction}\label{sec:intro}

Lipid molecules that comprise the cell membrane are free to flow within the 2D surface that represents the bilayer, but exchange momentum with the adjacent 3D medium. Insoluble surfactant monolayers and self-assembled polymer or nanoparticle layers have similar flow physics, driven by this unique `quasi-2D' nature of momentum transport. In their seminal work, \citet{Saffman1975} approximated the membrane as a thin Newtonian fluid layer sandwiched between and coupled to Stokes flow in adjacent 3D fluid phases. Proteins, ion channels, molecular motors, or synthetic particles are embedded in and constrained to move within this 2D layer. Such a description reveals a subtle transition from 2D to 3D hydrodynamics: the interface decouples from the bulk fluid in the `membrane-dominant' limit when surface viscous stresses far exceed the traction from the surrounding 3D fluid. This limit is described by the 2D Stokes equation and has no solution for the steady translation of a cylinder: a consequence of the Stokes paradox \citep{Manikantan2020JFM}. However, \citet{Saffman1976} recognized that subphase viscous stresses eventually catch up with surface viscous stresses beyond a critical length scale, ultimately regularizing the divergence inherent to 2D Stokes flow. Building on this framework, the hydrodynamics of disks \citep{Hughes1981,Stone1998}, spheres \citep{Danov2000,Fischer2006}, rods \citep{Levine2004,Fischer2004} and ellipsoids \citep{Stone2015} embedded in surface viscous interfaces are now firmly established, and widely employed in quantifying lateral diffusion in membranes, monolayers, and biofilms.

Most of these efforts, however, address flow around single particles. Correlated diffusion and collective effects become relevant in biological membranes that contain a high concentration of proteins \citep{Bussell1995,Oppenheimer2009}. Recent efforts have highlighted the role of long-ranged hydrodynamics in aggregating and assembling active \citep{Oppenheimer2019,Manikantan2020PRL} and driven \citep{Vig2023} membrane-bound point particles. While a point-particle description offers useful insight into this complex problem, real particles have a finite size and orientability with anisotropic hydrodynamic mobilities \citep{Levine2004,Fischer2004} that lead to nontrivial dynamics \citep{Camley2013,Shi2022}. Cellular processes like signaling, trafficking, and curvature sensing are associated with the transport and reorganization of filamentous proteins and rod-like domains \citep{Simunovic2013} on the plasma membrane. Recent advances also enable synthetic assembly of catalytic \citep{Dhar2006} and DNA origami-based \citep{Khmelinskaia2021} nanorods embedded in monolayers and membranes, yet no description of their collective surface viscous interactions yet exist. We aim to address this gap by borrowing fluid mechanical insights from analogous work on the dynamics of 3D suspensions of settling particles \citep{Koch1989}.

In \S \ref{sec:model} we develop a mean-field description of a dilute quasi-2D suspension of driven membrane-attached slender particles by coupling their anisotropic mobilities to long-ranged interfacial viscous hydrodynamics. In \S \ref{sec:stability}, we analyze the linear stability of such a system to concentration perturbations, revealing signatures of the resolution of the Stokes paradox in collective dynamics. In \S \ref{sec:mechanism}, we connect straining fields set up around driven particles to suspension stability and reveal mechanisms for a length scale selection in this problem. 

\section{Theoretical formulation}\label{sec:model}
\subsection {Mean-field description}%It is straightforward to generalize the following analysis to the case of 3D fluids on either side of the membrane. 

\begin{figure}
  \centerline{\includegraphics[width=0.75\textwidth]{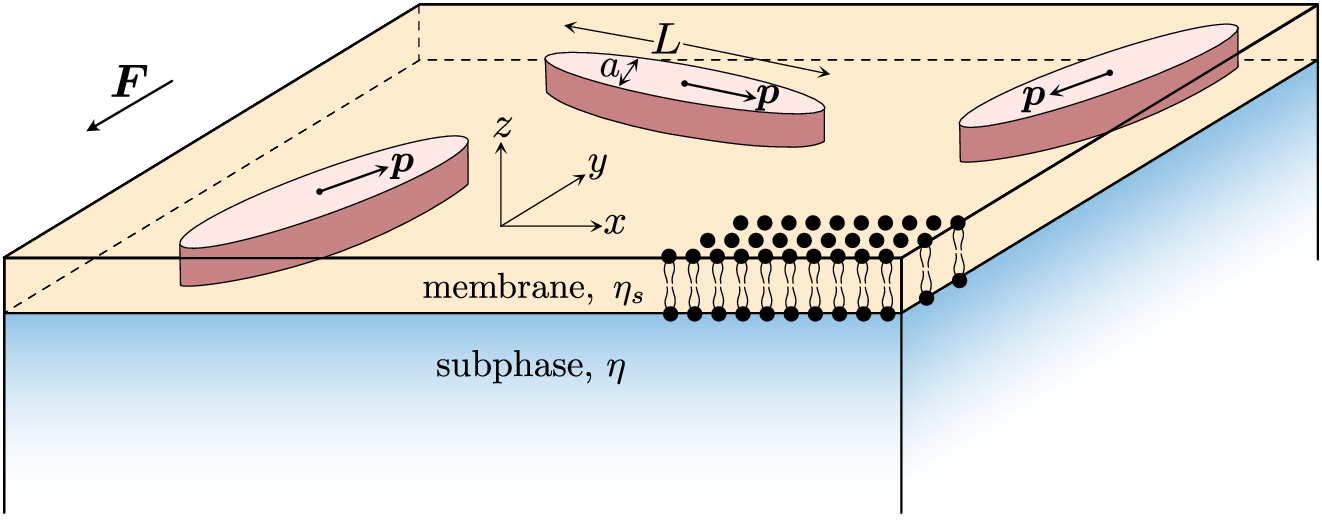}}
  \caption{System geometry: elongated particles ($L\gg a$) are embedded within an infinitesimally thin 2D viscous layer atop a 3D fluid subphase, and are all driven in the membrane plane by an external force $\bff$.}
\label{fig:sketch}
\end{figure}

The geometry of our system is shown in figure \ref{fig:sketch}. Slender rod-like particles of length $L$ and characteristic thickness $a$ (with $a \ll L$), each with a unit orientation vector $\bfp=(\cos\theta,\sin\theta)$, are embedded within a 2D viscous layer atop a 3D subphase. In a mean-field description, we define a probability distribution $\psi(\bfx,\bfp,t)$ such that the local concentration $c(\bfx,t)$ is obtained by integrating $\psi$ across all possible orientations $\bfp$:
 \begin{equation}
c(\bfx,t)= \int \psi (\bfx,\bfp,t) \intd\bfp=\int_0^{2\upi} \!\! \psi (\bfx,\bfp,t) \intd\theta.
 \end{equation}
We will also define $n$ as the number density of particles:
\begin{equation}
 n=\frac{1}{A} \int_A c(\bfx,t) \intd \bfx, 
 \end{equation} 
 where $A$ is the area of the membrane. Rods are constrained to translate and rotate in the plane of the membrane. Conservation of particles is then expressed by
 \begin{equation}\label{eq:cons}
 \pd{\psi}{t} + \bnabla_{\!s} \bcdot \left( \dot{\bfx} \psi \right) + \bnabla_{\!p} \bcdot \left( \dot{\bfp} \psi \right) = 0,
 \end{equation}
where $\dot{\bfx}$ and $\dot{\bfp}$ are translational and rotational velocities that capture probability flux. The surface gradient operator in the plane of the membrane is defined as $\bnabla_{\!s} =(\iden - \bfn\bfn)\bcdot \bnabla$ where $\bfn$ is the local normal to the 2D manifold that represents the membrane, and the orientational gradient operator simplifies to 
\begin{equation}
\bnabla_{\!p} = (\iden - \bfp\bfp)\bcdot \pd{}{\bfp} = \skew3\hat{\boldsymbol{\theta}} \pd{}{\theta}.
\end{equation}
In this initial work, we will restrict ourselves to planar membranes at $z=0$, and so $\bfn=\hat{\bfz}$.

\subsection {Micromechanical model}
We will explore the response of such a quasi-2D suspension of elongated particles to an externally applied force. For simplicity, the force $\bff$ will be taken to be a constant and acting in the same direction on all particles, akin to sedimentation in 3D suspensions \citep{Koch1989}. In membranes, this might represent driven or anchored collections of trans-membrane proteins in the context of cell motility \citep{Fogelson2014}. The translational velocity in \eqref{eq:cons} then has contributions from the self-mobility of each particle, and from advection due to the disturbance field generated by neighboring rods:
\begin{equation}
\dot{\bfx} = \bfu_{\!s}+\bfu_{\!d}.
\end{equation}
Here, $\bfu_{\!s}$ is the local response of a rod-like particle to a constant force $\bff$:
\begin{equation}\label{eq:us}
\bfu_{\!s} = \mu_\perp \left( \iden - \bfp \bfp\right) \bcdot \bff + \mu_\parallel \bfp\bfp\bcdot\bff,
\end{equation}
with hydrodynamic mobilities $\mu_\perp$ and $\mu_\parallel$ for translation in directions perpendicular and parallel to $\bfp$, respectively \citep{Fischer2004,Levine2004}.

Phospholipids that make up biological membranes and monolayers are insoluble and well approximated as 2D incompressible fluids \citep{Manikantan2020JFM}. The disturbance field $\bfu_{\!d}$ then solves the Boussinesq-Scriven equation \citep{Scriven1960} for the stress balance within a viscous incompressible interface of surface viscosity $\eta_s$ coupled to 3D Stokes flow in the adjacent bulk fluid of viscosity $\eta$ and forced by $\bff$ distributed at concentration $c(\bfx,t)$:
\begin{equation}\label{eq:bs}
\bnabla_{\!s} \mathit{\Pi} = \eta_s \nabla_{\!s}^2 \bfu_{\!d} - \eta \left. \frac{\partial \bfv}{\partial z} \right|_{z=0}  + \bff c(\bfx,t), \qquad \bnabla_{\!s} \bcdot \bfu_{\!d} = 0.
\end{equation}
Here, $\mathit{\Pi}$ is the surface pressure. Membrane flow is coupled to the 3D flow field $\bfv$ in the adjacent semi-infinite ($z$ from $0$ to $- \infty$) subphase via the traction term and a no-slip condition $\bfu=\bfv$ at $z=0$. This framework can be generalized to accommodate membrane curvature and to the presence of 3D fluids on either side of the membrane \citep{Shi2022,Shi2024}.

Finally, the rotational flux in \eqref{eq:cons} captures reorientation due to non-uniform disturbance fields. For slender rod-like particles, this is given by Jeffery's equation \citep{Jeffery1922}: 
\begin{equation}\label{eq:jeff}
\dot{\bfp} = (\iden -\bfp\bfp)\bcdot\bnabla_{\!s}\bfu_{\!d} \bcdot \bfp.
\end{equation} 
Thermal noise may contribute additional terms to $\dot{\bfx}$ and $\dot{\bfp}$, or equivalently as diffusive terms in the conservation equation \eqref{eq:cons}. We ignore diffusion in what follows to highlight the role of hydrodynamics alone: however, a diffusive flux is straightforward to add to the analysis below, and diffusion always acts to suppress large wave number fluctuations.

\section{Linear stability}\label{sec:stability}
We wish to examine the linear stability of a dilute quasi-2D suspension that is homogeneous in concentration and isotropic in orientational distribution. The disturbance field associated with such a base state is $\bfu_{\!d}=\boldsymbol{0}$. We perturb the probability distribution around this base state as $\psi(\bfx,\bfp,t) = n \left[ \psi_0 +\epsilon \psi'(\bfx,\bfp,t)  \right]$ where $\psi_0=(2\upi)^{-1}$, $\bfu_{\!d}=\epsilon \bfu'_{\!d}$ and  $\dot{\bfp}_{\!d}=\epsilon \dot{\bfp}'_{\!d}$. The corresponding local concentration is $c=n \left[1+\epsilon c' \right]$. Plugging perturbed quantities into \eqref{eq:cons} and recognizing the isotropic base state as well as 2D incompressibility of the membrane fluid field ($\bnabla_{\!s}\bcdot \bfu = 0$) gives the probability conservation equation at $O(\epsilon)$:
\begin{equation}\label{eq:cons_linearized}
\pd{\psi'}{t} + \bnabla_{\!s}\psi'\bcdot \bfu_{\!s} + \psi_0 \bnabla_{\!p}\bcdot \dot{\bfp}'_{\!d} = 0
\end{equation}
We then impose normal modes of the form
\begin{equation}\label{eq:normal}
\psi'(\bfx,\bfp,t) = \tilde{\psi}(\bfk,\bfp,\omega) \,e^{i (\bfk\bcdot \bfx - \omega t)},
\end{equation}
so that $\tilde{c}(\bfk,\omega)=\int \tilde{\psi} \intd\bfp$ is the associated local concentration. The associated hydrodynamic disturbance field can be obtained by solving \eqref{eq:bs} for the corresponding Fourier transform of the disturbance velocity \citep{Manikantan2020JFM}:
\begin{equation}\label{eq:uhat}
\tilde{\bfu}_d(\bfk,\bfp,\omega) = \frac{(\iden - \bfkhat\bfkhat)\bcdot n \bff \tilde{c}}{k^2 \eta_s+k \eta} ,
\end{equation}
where $k=|\bfk|$ and $\bfkhat=\bfk/k$ is a unit vector. Then, using Jeffery's equation \eqref{eq:jeff}, we find
\begin{equation}\label{eq:pdothat}
\tilde{\dot{\bfp}}_{\!d} = \frac{i (\bfp\bcdot\bfk) (\iden - \bfp\bfp) \bcdot (\iden - \bfkhat \bfkhat) \bcdot n \bff \tilde{c}}{k^2 \eta_s+k \eta} ~~\Rightarrow~~
\bnabla_{\!p}\bcdot \tilde{\dot{\bfp}}_{\!d} = \frac{-2i (\bfp\bcdot\bfk)  \bfp \bcdot (\iden - \bfkhat \bfkhat) \bcdot n \bff \tilde{c}}{k^2 \eta_s+k \eta}.
\end{equation}
Plugging \eqref{eq:normal} and \eqref{eq:pdothat} into \eqref{eq:cons_linearized} gives
\begin{equation}
\left(-i \omega + i \bfk \bcdot \bfu_{\!s} \right) \tilde{\psi}  - \frac{2i \psi_0 (\bfp\bcdot\bfk)  \bfp \bcdot (\iden - \bfkhat \bfkhat) \bcdot n \bff \tilde{c}}{k^2 \eta_s+k \eta} = 0,
\end{equation}
where $\bfu_{\!s}$ follows \eqref{eq:us}.

We now non-dimensionalize over characteristic values of $L$ for length, $F$ for force, and $F/(4\upi \eta_s)$ for velocities. We will take the direction of the force to be $\bffhat$ so that $\bff=F \bffhat$. Then,
\begin{equation}
\left(-\omega + \bfk \bcdot \left[\mu_\parallel\bfp\bfp + \mu_\perp (\iden - \bfp\bfp) \right] \bcdot \bffhat \right) \tilde{\psi}  - \frac{8 N\psi_0 \lsd (\bfp\bcdot\bfk)  \bfp \bcdot (\iden - \bfkhat \bfkhat) \bcdot \bffhat \tilde{c}}{k^2 \lsd+k } = 0,
\end{equation}
where $\lsd=\eta_s/\eta L$ is the dimensionless Saffman-Delbr{\"{u}}ck length. Alternatively, $\lsd$ can be interpreted as a Boussinesq number \citep{Manikantan2020JFM} which characterizes surface viscous stresses relative to 3D viscous stresses from the adjacent subphase fluid over the length $L$. Momentum transport is membrane-dominated when $\lsd\rightarrow \infty$ and subphase-dominated when $\lsd\rightarrow 0$. $N=n\upi L^2$ is an effective area fraction occupied by particles. The long-ranged description is valid only in the dilute regime, and so $N$ is at most $O(1)$.

To proceed, we simplify our problem by considering only perturbations in concentration along the direction perpendicular to the external forces: $\bfk \bcdot \bffhat = 0$. This is motivated by the fact that these transverse perturbations are the most unstable in analogous 3D sedimentation problems \citep{Koch1989} and driven 2D suspensions \citep{Vig2023}. Simplifying with this assumption and rearranging gives
\begin{equation}\label{eq:disp_temp}
\tilde{\psi}  = \left(\frac{8 N\psi_0 \lsd \tilde{c}}{k^2 \lsd+k } \right) \frac{(\bfp\bcdot\bfk) (\bfp\bcdot\bffhat) }{(\mu_\parallel - \mu_\perp)(\bfp\bcdot\bfk) (\bfp\bcdot\bffhat) -\omega} .
\end{equation}
Integrating \eqref{eq:disp_temp} across all orientations while noting that $\tilde{c}(\bfk,\omega)=\int \tilde{\psi} \intd\bfp$  eliminates $\tilde{\psi}$ to reveal an implicit relationship for $\omega(k)$:
\begin{equation}\label{eq:disp}
\int \!\! \frac{(\bfp\bcdot\bfk) (\bfp\bcdot\bffhat) }{(\mu_\parallel - \mu_\perp)(\bfp\bcdot\bfk) (\bfp\bcdot\bffhat) -\omega}  \intd \bfp = \frac{k^2 \lsd+k }{8 N \psi_0 \lsd }.
\end{equation}
The mobility coefficients $\mu_\parallel$ and $\mu_\perp$ are functions of $\lsd$, and $\psi_0=(2\upi)^{-1}$ is the isotropic base state distribution. For a given area fraction $N$, \eqref{eq:disp} thus represents a family of dispersion relations parametrized by the Saffman-Delbr{\"{u}}ck length $\lsd$.

\begin{figure}
  \centerline{\includegraphics[width=0.86\textwidth]{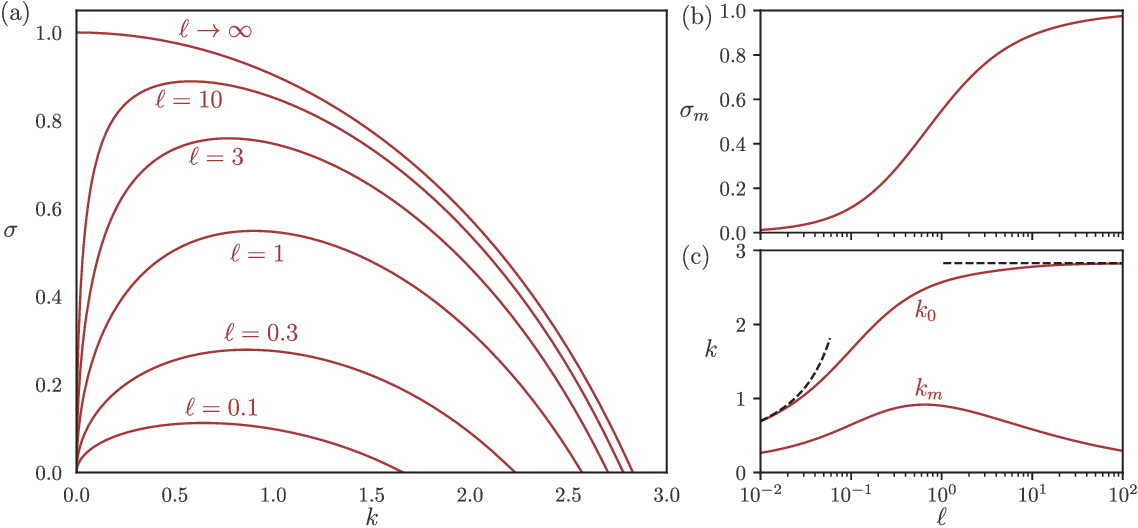}}
  \caption{(a) Growth rates $\sigma$ of unstable modes at various $\ell=\eta_s/\eta L$ for a dimensionless number density of $N=n \upi L^2=1$. (b) Maximum growth rate $\sigma_m$, and (c) most unstable wave number $k_m$ and largest wave number $k_0$ as a function of $\ell$. Dashed lines in (c) are asymptotic limits from \eqref{eq:maxklimits}.} 
\label{fig:dispersion}
\end{figure}

%Note again that the perturbation is assumed to be along the direction perpendicular to the forcing: $\bfk \bcdot \bffhat = 0$. 
In a 2D Cartesian plane representing the membrane, we can take $\bffhat=-\hat{\boldsymbol{y}}$ and $\bfkhat=\hat{\boldsymbol{x}}$ without loss of generality. Simplifying \eqref{eq:disp} using $\bfp=(\cos\theta,\sin\theta)$ then gives
\begin{equation}\label{eq:dispersiontheta}
\frac{1}{2\upi}\int_0^{2\upi} \!\! \frac{k \sin \theta \cos \theta }{(\mu_\parallel - \mu_\perp)k \sin \theta \cos \theta +\omega}  \intd \theta = \frac{k^2 \lsd+k }{8 N \lsd },
\end{equation}
which is an implicit integral equation for the dispersion relation $k(\omega)$. For normal modes of the form $\exp[i (\bfk\bcdot \bfx - \omega t)]$, perturbations are unstable if the imaginary part of $\omega$ is positive. 

Figure \ref{fig:dispersion}(a) shows the growth rate $\sigma=\Imag[\omega]$ as a function of wave number obtained by numerically evaluating roots of \eqref{eq:dispersiontheta} with a secant method. The $\lsd\rightarrow\infty$ limit is reminiscent of analogous 3D suspensions, where the system-spanning mode corresponding to $k=0$ is the most unstable \citep[see][]{Koch1989,Manikantan2014}. To obtain the growth rate at $k=0$ analytically, we expand the integrand in \eqref{eq:dispersiontheta} as a Taylor series in $k$: 
\begin{equation}
\frac{k}{2\upi \omega}\int_0^{2\upi} \!\!  \sin\theta\cos\theta \left[1- \frac{(\mu_\parallel - \mu_\perp)k}{\omega} \sin\theta\cos\theta + \dots \right] \intd \theta =\frac{k^2 \lsd+k }{8 N \lsd }.
\end{equation}
Integrating term by term and simplifying gives
\begin{equation}\label{eq:series}
-\frac{(\mu_\parallel - \mu_\perp) k^2}{\omega^2} + O(k^4)= \frac{k^2 \lsd + k}{N\lsd}.
\end{equation}
The dimensionless mobility difference is $\mu_\parallel - \mu_\perp\rightarrow 1$ \citep{Fischer2004,Levine2004} in the membrane-dominated regime ($\lsd \rightarrow \infty$), which gives $\omega^2=-N+O(k^2)$ or 
\begin{equation}\label{eq:sigmmax}
\sigma (k\rightarrow 0, \lsd\rightarrow \infty) = \sqrt{N}.
\end{equation}
In fact, this turns out to be the maximum value of the growth rate across all $k$ and $\lsd$. 

Increasing subphase viscous stresses (decreasing $\lsd$) always acts to suppress the instability. Notably, the $k=0$ mode is stable for finite $\eta$, reflecting a boundary layer at infinity that regularizes the singularity in 2D Stokes by accounting for subphase viscous stresses \citep{Saffman1976}. For finite $\lsd$, \eqref{eq:series} gives $\omega^2 \sim - k N \lsd (\mu_\parallel - \mu_\perp)$ or $\sigma=\Imag[\omega] \propto \sqrt{k}$. In other words, the growth rate goes to zero as $k\rightarrow 0$ for all values of $\lsd$ except in the limit $\lsd\rightarrow \infty$. The most unstable mode then corresponds to a finite value of $k$ which suggests a mechanism for wave number selection. The most unstable wave number $k_m$ can be determined by taking the derivative of \eqref{eq:dispersiontheta} with respect to $k$ and numerically solving the implicit integral equation corresponding to $\partial \omega/\partial k = 0$. 
% \begin{equation}
% \int_0^{2\upi} \!\! \left[ \frac{ \sin \theta \cos \theta }{(\mu_\parallel - \mu_\perp)k_m \sin \theta \cos \theta +\omega (k_m)} \right]^2 \!\! \intd \theta = - \frac{\pi }{4 N (\mu_\parallel - \mu_\perp)}.
% \end{equation}
% where $\omega$ solves \eqref{eq:dispersiontheta}. We solve this implicit equation for $k_m$ using numerical quadrature and a secant root-finding method. 
Figure~\ref{fig:dispersion}(b,c) shows $k_m$ and the corresponding growth rate $\sigma_m$ as a function of $\lsd$. Note that the maximum growth rate $\sigma_m$ asymptotes to the $\lsd\rightarrow\infty$ limit as predicted by \eqref{eq:sigmmax}, and monotonically decreases upon decreasing $\lsd$. The wave number $k_m$ corresponding to fastest growth of perturbations, however, varies non-monotonically with $\lsd$ and peaks at $\lsd=O(1)$: in \S\ref{sec:mechanism} we propose a mechanism of instability that clarifies the length scale selection underlying $k_m$ and its dependence on $\lsd$.
 
Finally, a key parameter is the largest unstable wave number $k_0$ which reveals the length scale of shortest unstable modes. Setting $\omega=0$ in \eqref{eq:dispersiontheta} gives
\begin{equation}
k_0^2 \lsd+k_0=\frac{8 N \lsd}{\mu_\parallel - \mu_\perp}.
\end{equation}
Solving this quadratic equation for positive $k_0$ yields
%\begin{equation}\label{eq:maxk}
%k_0 = \frac{-1+ \sqrt{1+\frac{32 N \lsd^2}{\mu_\parallel - \mu_\perp}}}{2\lsd},
%\end{equation}
the curve in figure~\ref{fig:dispersion}(c). Note that this $\omega=0$ limit is only valid for $k\neq 0$. The range of unstable modes increases monotonically with $\lsd$. The high and low $\lsd$ limits can be found analytically: $\mu_\parallel - \mu_\perp$ asymptotes to $1$ and $4\lsd \left[ \ln (0.48 \lsd)-1 \right]$ in these limits respectively \citep{Fischer2004,Levine2004} giving
\begin{equation}\label{eq:maxklimits}
k_0 (\lsd \rightarrow \infty) = \sqrt{8N}, \qquad k_0 (\lsd \rightarrow 0)= \frac{2N}{-\log\left(\lsd \right)-1.73}.
\end{equation}
These are shown as the dashed curves in figure~\ref{fig:dispersion}(c).

\section{Instability mechanism and length scale selection}\label{sec:mechanism}

\begin{figure}
  \centerline{\includegraphics{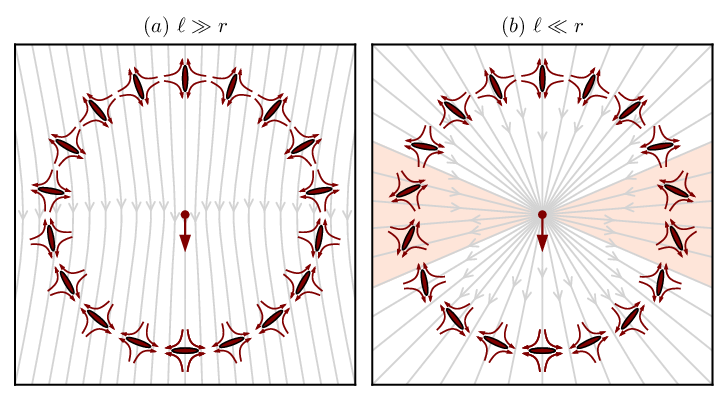}}
  \caption{Mechanism of the instability in terms of the disturbance due to a point-force (at the center) on the orientation of elongated particles at a distance $r$. Gray streamlines represent the in-plane velocity $\bfu$. The magenta extensional streamlines around each neighboring particle depict the symmetric part of the local, linear shear field. For $\ell \gg r$ (membrane-dominated regime), neighbors always reorient such that they are drawn towards the point force, thereby increasing particle density. By contrast, at much longer length scales $r \gg \ell$ (subphase-dominated regime), particles that fall within the shaded region are drawn away from the point force, reducing particle density and stabilizing the suspension. For a fixed $\eta_s$ and $\eta$, the system transitions continuously from membrane-dominated to subphase-dominated at large enough distances. }
\label{fig:mechanism}
\end{figure}

The mechanism of instability can be understood by examining the disturbance flow created in the plane of the membrane by a point force corresponding to a reference particle, and the effect of this flow on realigning neighboring particles. As all particles are driven in the same direction, and since alignment dictates direction of mobility given a force, such a picture helps identify microstructural configurations that favor aggregation or separation. 

The dimensionless velocity at a point $\bfr$ in the plane of the membrane due to a point force at the origin is \citep{Fischer2004,Levine2004,Manikantan2020JFM}
\begin{equation}\label{eq:vel_field}
\begin{aligned}
\bfu (\bfr) &=  \upi \left[  \left( \frac{H_1(d)}{d} -\frac{2}{\upi d^2} - \frac{Y_0(d)+Y_2(d)}{2} \right) \frac{\bfr \bfr}{r^2}+ \right. \\ 
&+ \left. \left(H_0(d)- \frac{H_1(d)}{d}  +\frac{2}{\upi d^2} - \frac{Y_0(d)-Y_2(d)}{2} \right) \left( \iden - \frac{\bfr \bfr}{r^2}  \right) \right]\bcdot \bffhat,
\end{aligned}
\end{equation}
where $r=|\bfr|$ is the dimensionless distance, and $d=r/\lsd$. Here, $H_\nu$ and $Y_\nu$ are Struve and Bessel functions of the second kind or order $\nu$, respectively. Note that $\lsd$ is renormalized by the distance $r$ in the far-field description: in other words, at constant $\eta_s$ and $\eta$, the system switches from membrane-dominated to subphase-dominated at large enough length scales. 

The velocity field in the plane of the membrane from \eqref{eq:vel_field} asymptotes to that corresponding to the 2D stokeslet in the membrane-dominated regime ($\lsd\gg r$):
\begin{equation}\label{eq:binf_limit}
\bfu (\bfr,\lsd\gg r) \rightarrow \left[ -\log\left( \frac{r}{\lsd}\right)\iden + \frac{\bfr \bfr}{r^2} \right] \bcdot \bffhat.
\end{equation}
The mechanism of the instability is tied to the reorientation of neighboring particles when placed in such a disturbance field. Specifically, rod-like particles align with the extensional axis of the local flow. The rate-of-strain tensor in the membrane-dominated case is
\begin{equation}
\mathsfbi{E} (\lsd \gg r ) = \frac{\bnabla_{\!s} \bfu_{\!d} + \bnabla_{\!s} \bfu_{\!d}^T}{2} = \frac{(\bfr \bcdot \bffhat) \iden }{r^2} - \frac{ 2 \bfr \bfr \bfr \bcdot \bffhat}{r^4} .
\end{equation}
Using $\bffhat=-\hat{\boldsymbol{y}}$ and $\bfr=r \hat{\boldsymbol{r}} = (r \cos \phi, r\sin \phi)$ where $\phi$ is measured relative to the positive $x$ direction diagonalizes the rate-of-strain tensor in the $(\hat{\boldsymbol{r}},\phihat)$ basis as
\begin{equation}\label{eq:eigen_highbous}
\mathsfbi{E} (\lsd\gg r ) = \frac{\sin \phi}{r}
\begin{bmatrix}
1 & 0 \\
0 & -1 
\end{bmatrix} .
\end{equation}
The eigenvectors of $\mathsfbi{E}$ are the principal directions of extension. We see from \eqref{eq:eigen_highbous} that the extensional axis is along $\hat{\boldsymbol{r}}$ (when $\phi>0$) or perpendicular to $\hat{\boldsymbol{r}}$ (when $\phi<0$) as shown in figure~\ref{fig:mechanism}(a). Elongated particles thus always align in a direction that draws them towards the reference particle. This is analogous to the classic 3-D instability in a  suspension of rod \citep[see][]{Koch1989}. Note also from \eqref{eq:eigen_highbous} that the eigenvalues of $\mathsfbi{E}(\lsd \gg r)$, corresponding to the local rates of extension $\lambda_{\rm ext}$, peak at $\phi=\pm \shalf\upi$ and vanish at $\phi=0$: rods on the same horizontal line thus do not reorient in response to the disturbance due to the other. 

\begin{figure}
  \centerline{\includegraphics[width=\textwidth]{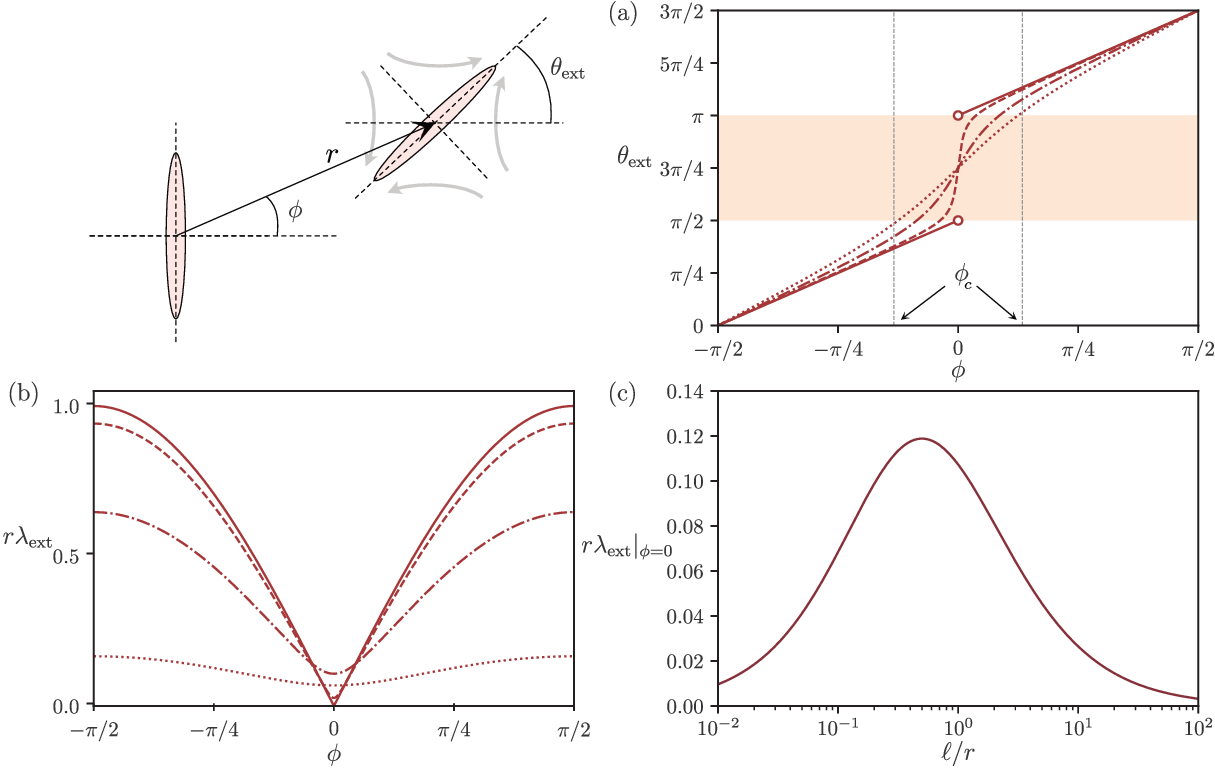}}
  \caption{(a) Principal axes of stretch $\theta_{\rm ext}$ corresponding to preferred orientation of a neighboring particle as a function of azimuthal position $\phi$ around a reference particle for $\lsd/r=0.1$ (dotted line), $1$ (dash-dot), and $10$ (dashed line). The shaded patch denotes the set of orientations that draws particles away from each other: this region is not accessed in the limit of $\lsd \rightarrow \infty$ (solid line), whereas $\lsd\rightarrow 0$ maximizes this window for a range of relative positions $|\phi|<|\phi_c|=\sin^{-1}(1/\sqrt{6})$. (b) Principal rates of stretch $\lambda_{\rm ext}$ corresponding to cases shown in (b). (c) $\lambda_{\rm ext}$ does not vanish at $\phi=0$ for finite $\lsd$, and peaks when $r\sim\lsd$. }
\label{fig:eigen}
\end{figure}

The in-plane extensional field surrounding a point force changes dramatically upon increasing the contribution from the subphase (decreasing $\lsd$). The membrane velocity field in the surface-inviscid limit ($\lsd\ll r$) is
\begin{equation}\label{eq:b0_limit}
\bfu (\bfr,\lsd \ll r ) \rightarrow 2 \lsd ~ \frac{\bfr \bfr}{r^3} \bcdot \bffhat.
\end{equation}
Note that \eqref{eq:b0_limit} is not simply a 2-D slice of the flow due to a point force in the bulk 3-D fluid: surface incompressibility requires that $\bfu$ be divergence-free in the plane of the membrane. The rate-of-strain tensor is no longer easily diagonalizable in the $(\hat{\boldsymbol{r}},\phihat)$ basis; however, $\mathsfbi{E}$ still reveals key features when expressed in the $(\hat{\boldsymbol{x}},\hat{\boldsymbol{y}})$ basis:
\begin{equation}\label{eq:eigen_lowbous}
\mathsfbi{E} (\lsd \ll r ) = \frac{2 \lsd}{r^2}
\begin{bmatrix}
\sin \phi (3 \cos^2 \phi -1) & \cos\phi(3 \sin^2\phi-\shalf) \\
\cos\phi(3 \sin^2\phi-\shalf) & \sin\phi(3\sin^2\phi-2)
\end{bmatrix} .
\end{equation}
First, principal rates of extension scale with $\lsd$, consistent with a weakening instability as seen in \S\ref{sec:stability}. Additionally, neighboring particles along the wave vector (on the same horizontal line as the point force, or along $\phi=0$) experience a non-zero rate of extension along a principal direction $\theta_{\rm ext}=3\upi/4$ (figure~\ref{fig:eigen}). Coupled to the anisotropic mobility of rods, local extension therefore reorients rods in a manner that draws them away from the reference particle if initially placed in a region near $\phi=0$, which acts as a stabilizing mechanism. Note also from \eqref{eq:eigen_lowbous} that  $\mathsfbi{E}$ is diagonal in the $(\hat{\boldsymbol{x}},\hat{\boldsymbol{y}})$ basis when $3 \sin^2\phi=\shalf$: i.e., for a critical angle $\phi_c=\pm \sin^{-1}(1/\sqrt{6})$. This corresponds to principal axes that are aligned with $\hat{\boldsymbol{x}}$ or $\hat{\boldsymbol{y}}$ and reveals a transition from configurations that draw neighboring particles towards the point force (when $|\phi|>|\phi_c|$) or away from it. This also suggests a mechanism for wave number selection as seen in the stability analysis in \S\ref{sec:stability}: the critical length scale emerges from a balance of fluxes of particles driven out from within the region $|\phi|<|\phi_c|$ versus drawn in from regions where $|\phi|>|\phi_c|$.

Principal directions of extension for arbitrary $\lsd$, obtained by numerically evaluating the eigenvectors of $\mathsfbi{E}$ from the general velocity field  in \eqref{eq:vel_field}, are shown in figure~\ref{fig:eigen}(a). This further clarifies the emergence of a `stabilizing region' in terms of preferred orientations for finite $\lsd$. Particles placed in this region are drawn away from the reference particle, contributing to a suppression of the instability.  The $\lsd/r \rightarrow \infty$ limit avoids this region entirely as seen mechanistically in figure~\ref{fig:mechanism} and quantitatively as a discontinuity in $\theta_{\rm ext}(\phi)$ in figure~\ref{fig:eigen}(a). This reflects the singularity of 2D stokes flow at infinity, and the subphase viscous contribution introduces a boundary layer that regularizes this discontinuity. In doing so, $\theta_{\rm ext}$ takes on values that fall within the shaded patch in figure \ref{fig:eigen}(a) that represents a stabilizing region, where particle orientations are such that they are drawn away from the reference particle. This mechanism also explains the stabilization of the $k=0$ mode for finite $\lsd $ in \S\ref{sec:stability}: for any $\eta$ and $\eta_s$ (and since $\bfk$ is parallel to $\hat{\boldsymbol{x}}$ for transverse perturbations), particles along the horizontal line ($\phi=0$) will always fall in the stabilizing region at a large enough value of $r$.

The associated eigenvalues of $\mathsfbi{E}$ are local rates of extension $\lambda_{\rm ext}$: while these principal rates of extension always peak at $\phi=\pm \shalf\upi$ and their magnitudes decrease upon decreasing $\lsd$, they do not always vanish at $\phi=0$ (figure~\ref{fig:eigen}b,c). Indeed, $\lambda_{\rm ext}(r,\phi=0)$ peaks when $r\sim\lsd$. For a pair of particles placed in the same horizontal line, the reorientation that favors particle separation is thus strongest at distances comparable to the Saffman-Delbr{\"{u}}ck length. 

\section{Conclusion}
In this paper, we developed a mean-field model to examine hydrodynamic collective modes of elongated particles embedded in viscous membranes or monolayers. We focused on a linear stability analysis around a homogeneous and isotropic `quasi 2D suspension' of such particles. We discover a unique mechanism for stabilization of such a system, reflecting aspects of the Stokes paradox in 2D viscous flows and its regularization via 3D subphase stresses. We then tie this mechanism to the interplay between anisotropic particle mobility and long-ranged hydrodynamics in the plane of the membrane. 

More generally, this mean-field framework opens a rich avenue of fluid dynamical problems and tools to examine microstructure on biological or synthetic membranes. The method can be readily adapted to membranes with curvature using modified Green's functions \citep{Shi2022,Shi2024}, to active particles on membranes via a stresslet solution \citep{Manikantan2020PRL,Oppenheimer2019}, and to systems with weakly non-Newtonian membrane rheology using tools such as the Lorentz reciprocal theorem \citep{Vig2023}. We have illustrated the `deep' subphase limit: confining the 3D fluid has been shown to amplify bulk stresses \citep{Camley2013,Shi2024} and modify membrane flow fields in manners that favor aggregation \citep{Manikantan2020PRL}. This modification is also straightforward within the mean-field description developed here. Our findings also open up opportunities to examine long-term nonlinear dynamics in these systems by computationally solving the mean-field model and through efficient particle simulations of crowded systems. Building on these insights, we anticipate that the present work will spur new directions of fluid dynamical studies into active and passive quasi-2D suspensions. 

\backsection[Funding]{This material is based upon work supported by the U.S. National Science Foundation under Grant No. CBET-2340415.}

\backsection[Declaration of interests]{The author reports no conflict of interest.}

\backsection[Author ORCIDs]{H. Manikantan, https://orcid.org/0000-0002-4270-3527}

\bibliographystyle{jfm}
%\bibliography{meanfield_refs}

\end{document}